\documentclass[12pt]{article}
\usepackage[margin=1in]{geometry}
\usepackage{amsmath, amsfonts}
\usepackage{graphicx}
\usepackage{natbib}
\usepackage{url}
\usepackage{sectsty}
\usepackage{pgfplots}
\usepackage{pdfpages}
\usepackage{mathtools}
\usepackage{setspace}
\usepackage{authblk}
\usepackage[defaultcolor=black]{changes}
\usepackage{float}

\mathtoolsset{showonlyrefs}

\defcitealias{wpp2022}{United Nations, 2022b}
\defcitealias{un2022wpp}{United Nations, 2022a}
\defcitealias{expert2018international}{EGRIS, 2018}
\defcitealias{jetson2024}{UNHCR 2024a}
\defcitealias{jetson2024somalia}{UNHCR 2024b}
\defcitealias{ussd2007}{U.S. Department of State 2007}
\defcitealias{un1951convention}{UN 1951}
\defcitealias{un1967convention}{UN 1967}

\newcommand{\arxiv}{1}

\ifx\arxiv\undefined 
    \date{}
    \author{}
\else
    \author[1]{Herbert P. Susmann}
    \author[2]{Adrian E. Raftery}
    \date{}
    
    \affil[1]{\small Division of Biostatistics, Department of Population
      Health, NYU Grossman School of Medicine, USA. Email: susmah01@nyu.edu (Corresponding author)}
    
    \affil[2]{\small Departments of Statistics and Sociology, University of Washington, USA. Email: raftery@uw.edu}
\fi

\newcommand{\pipelinefigure}{
\begin{figure}[!ht]
    \centering
    \includegraphics[width=\columnwidth]{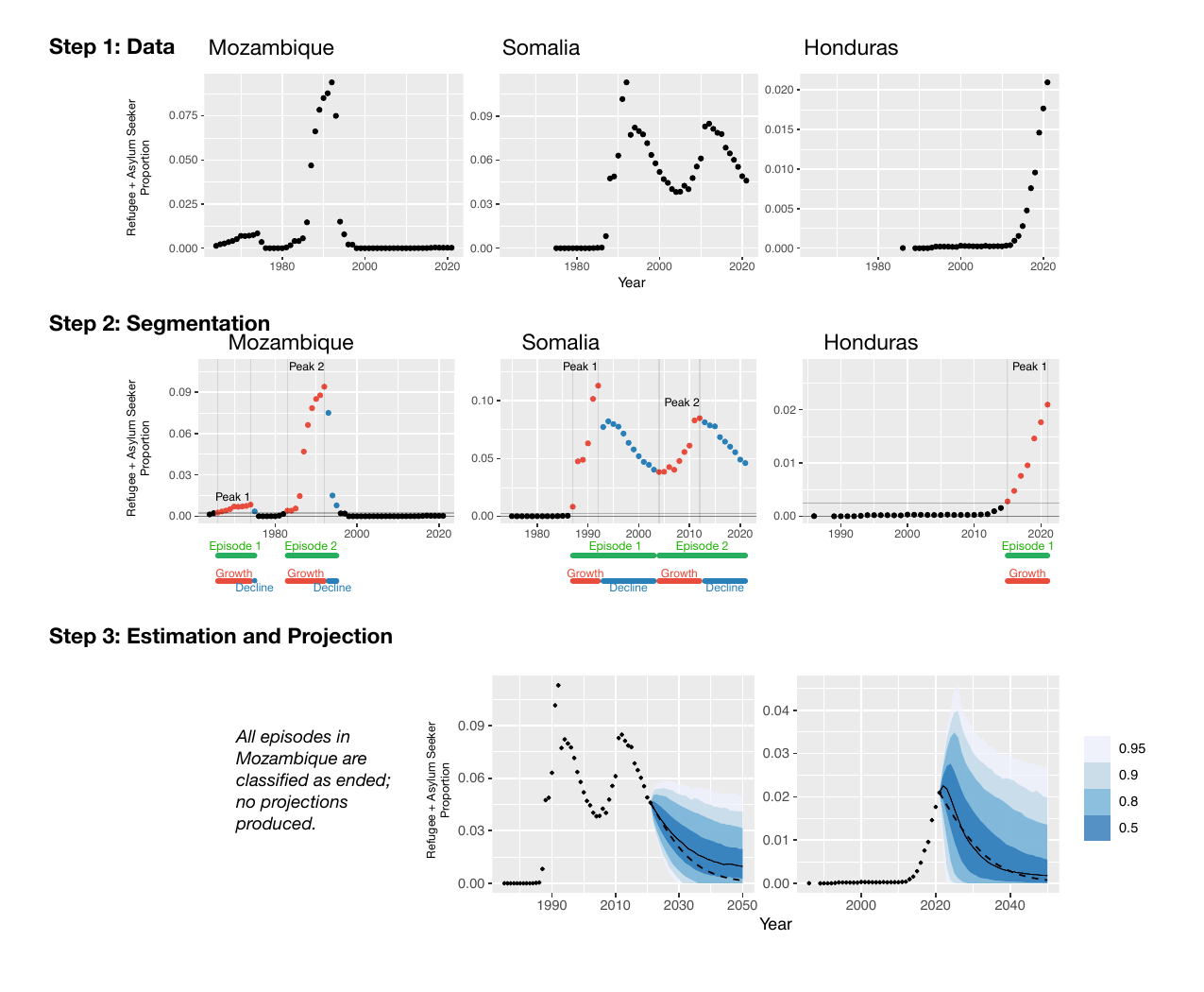}
    \caption{Overall modeling pipeline for projecting refugee and asylum seeker populations. Step 1 illustrates the observed data from Mozambique, Somalia, and Honduras. In Step 2, the data are segmented into multiple refugee crises. In Step 3, projections are produced for ongoing crises.}
    \label{fig:pipeline}
\end{figure}
}

\newcommand{\logisticmodelfigure}{
\begin{figure}[!ht]
    \centering
    \includegraphics[width=0.6\columnwidth]{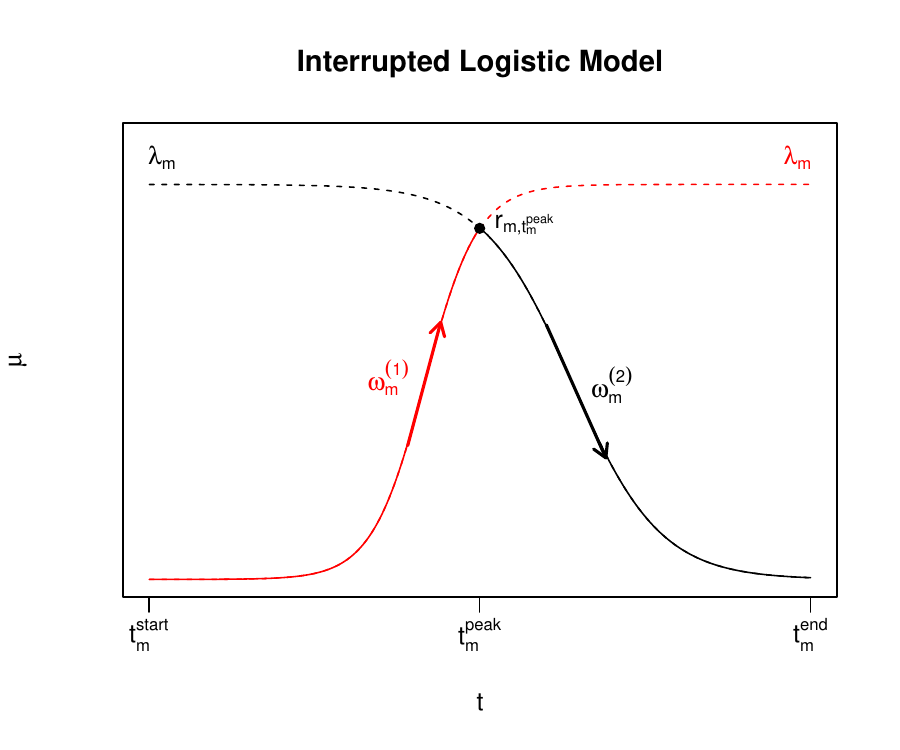}
    \caption{Diagram of the interrupted logistic process model.}
    \label{fig:logistic_model}
\end{figure}
}

\newcommand{\interruptedlogisticongoingfigure}{
\begin{figure}[!ht]
    \centering
    \includegraphics[width=0.7\columnwidth]{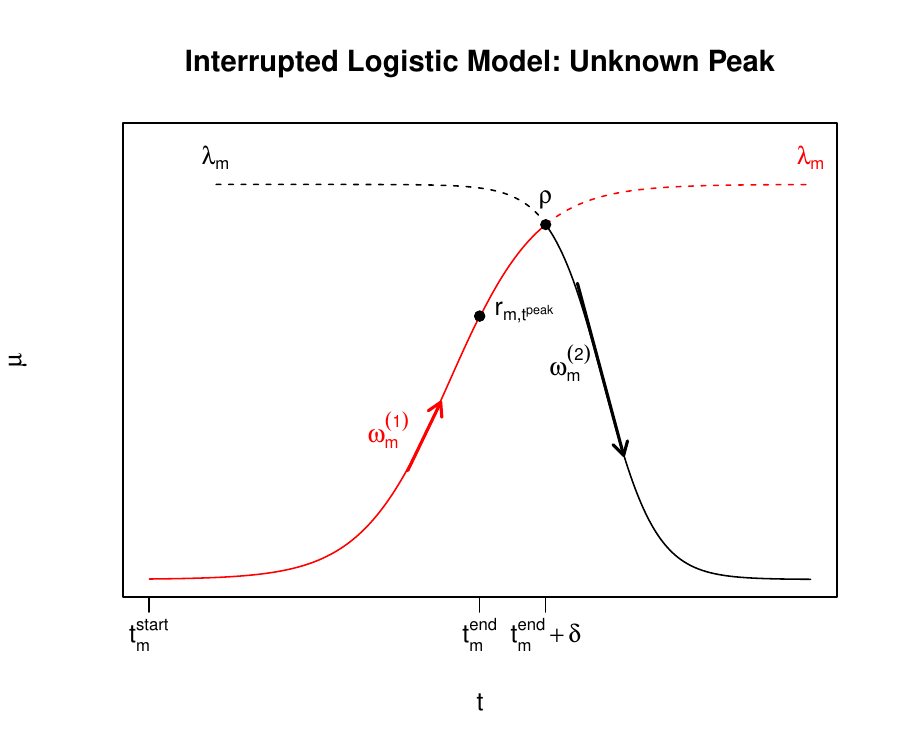}
    \caption{Diagram of projections for ongoing \added{episode} in the growth phase. Projections are generated conditional on a parameter $\delta$ that controls how long the \added{episode} will continue before reaching its peak.}
    \label{fig:interrupted-logistic-ongoing}
\end{figure}
}

\newcommand{\exampleprojectionsfigure}{
\begin{figure}[!ht]
    \centering
    \includegraphics[width=0.9\columnwidth]{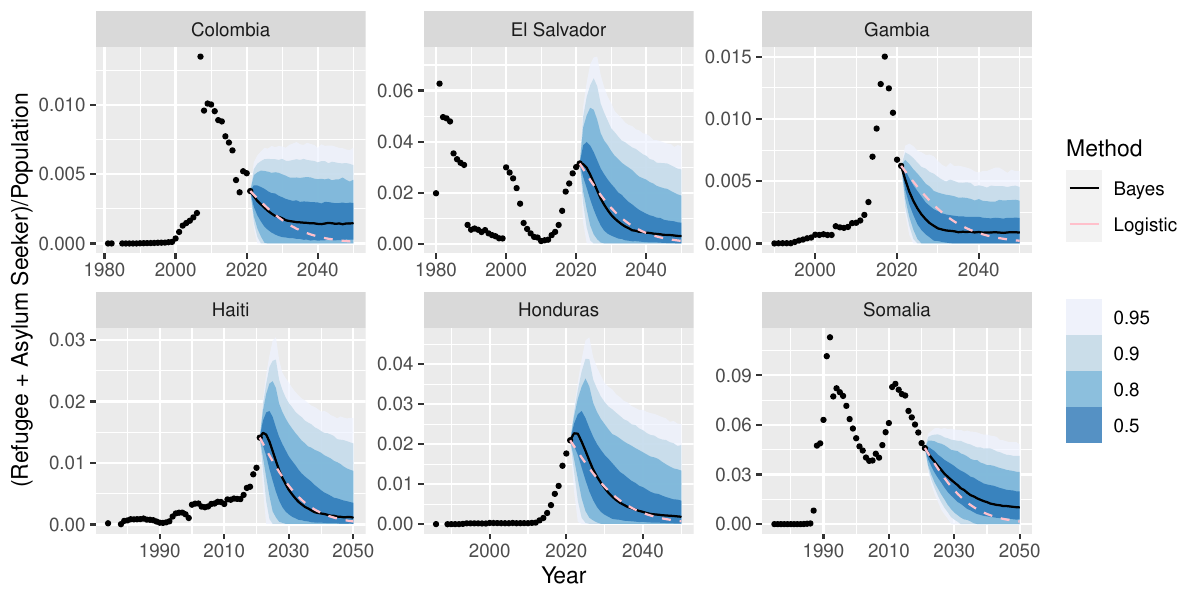}
    \caption{Projections for six illustrative countries. Shaded bands show the 50\%, 80\%, 90\%, and 95\% credible intervals. The pink dotted line shows the benchmark deterministic projection method.}
    \label{fig:example-projections}
\end{figure}
}

\newcommand{\exampleprojectionfigure}{
\begin{figure}[!ht]
    \centering
    \includegraphics[width=0.9\columnwidth]{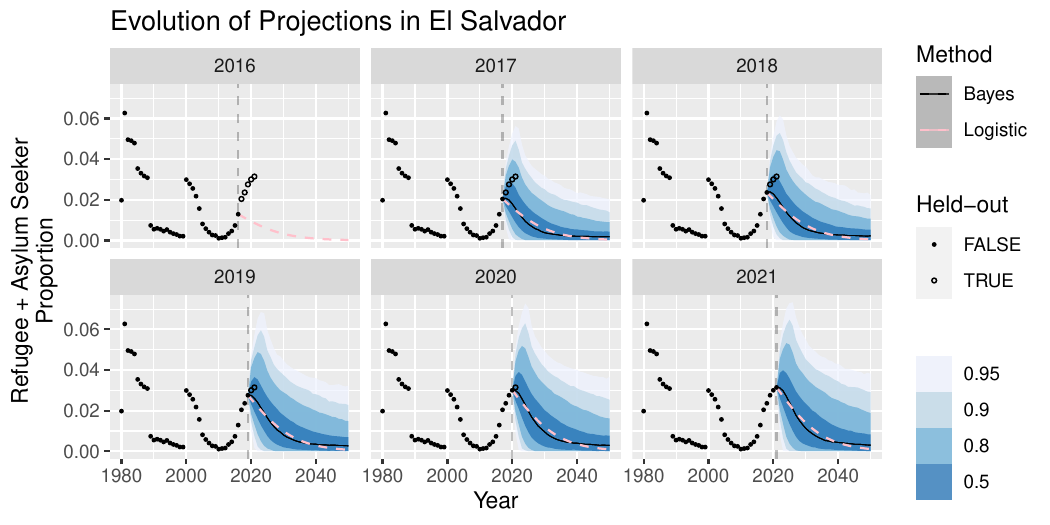}
    \caption{Projections in El Salvador from 2016-2021. Shaded bands show the 50\%, 80\%, 90\%, and 95\% credible intervals, respectively. The pink dotted line shows the benchmark deterministic projection method.}
    \label{fig:projection-el-salvador}
\end{figure}
}
\newcommand{\classificationtable}{
\begin{table}[]
    \centering
    \caption{Countries in the analysis dataset grouped by classification of their most recent \added{episode}. \added{Country names are presented as in the data source.}}
    \label{tab:country_crisis_classifications}
\
    \begin{tabular}{|p{0.95\columnwidth}|}
    \hline
    \textbf{Ended episodes} \\
    Angola, Cambodia, Chad, Equatorial Guinea, Ethiopia, Guinea-Bissau, Lao People's Dem. Rep., Liberia, Mozambique, Namibia, Sierra Leone, Slovenia, Tajikistan, Timor-Leste, Togo, Uganda, Zimbabwe \\
    \\
    \textbf{Ongoing episodes, decline phase} \\
    Afghanistan, Azerbaijan, Bhutan, Burundi, Colombia, Croatia, Gambia, Rwanda, Serbia and Kosovo: S/RES/1244 (1999), Somalia \\
    \\
    \textbf{Ongoing episodes, growth phase} \\
    Albania, Armenia, Bosnia and Herzegovina, Central African Rep., Congo, Dem. Rep. of the Congo, Djibouti, El Salvador, Eritrea, Georgia, Guatemala, Guinea, Guinea-Bissau, Haiti, Honduras, Iraq, Lebanon, Mali, Mauritania, Myanmar, Nicaragua, South Sudan, Sri Lanka, Sudan, Syrian Arab Rep., Venezuela (Bolivarian Republic of)
    \\
    \hline
    \end{tabular}
\end{table}
}

\newcommand{\errortable}{
\begin{table}[]
    \centering
    \caption{\label{tab:validation_error} Mean Absolute Error (MAE) and Mean Error (ME) of posterior median projections in the validation exercises.}
    \begin{tabular}{|llrrrrr|}
    \hline
    & & & \multicolumn{2}{c}{MAE} & \multicolumn{2}{c|}{ME} \\
    Cutoff & Target & $n$ & Bayes & Logistic & Bayes & Logistic \\
    \hline
    \multicolumn{7}{|l|}{\textit{1 year ahead}} \\
    2016 & 2017 & 25 & \textbf{0.73} & 0.82 & \textbf{0.62} & 0.75\\
    2017 & 2018 & 26 & \textbf{0.20} & 0.28 & \textbf{0.05} & 0.23\\
    2018 & 2019 & 26 & \textbf{0.13} & 0.20 & \textbf{-0.03} & 0.18\\
    2019 & 2020 & 28 & \textbf{0.15} & 0.27 & \textbf{0.01} & 0.22\\
    2020 & 2021 & 29 & \textbf{0.26} & 0.43 & \textbf{0.20} & 0.40\\
    \multicolumn{7}{|l|}{\textit{5 year ahead}} \\
    2011 & 2016 & 16 & \textbf{1.46} & 1.58 & 0.89 & \textbf{0.79}\\
    2016 & 2021 & 22 & 1.85 & \textbf{1.76} & \textbf{1.67} & 1.70\\
    \multicolumn{7}{|l|}{\textit{10 year ahead}} \\
    2011 & 2021 & 13 & \textbf{2.26} & 2.28 & 1.88 & \textbf{1.77}\\
    \hline
    \end{tabular}
\end{table}
}

\newcommand{\coveragetable}{
\begin{table}[]
    \centering
    \caption{Empirical coverage of 80\%, 90\%, and 95\% credible intervals for projections in the validation exercises.}
    \label{tab:validation_coverage}
    \begin{tabular}{|llrrrr|}
    \hline
    & & & \multicolumn{3}{c|}{Coverage} \\
    Cutoff & Target & $n$ & 80\% & 90\% & 95\% \\
    \hline
    \multicolumn{6}{|l|}{\textit{1 year ahead}} \\
    2016 & 2017 & 25 & 88.0\% & 88.0\% & 88.0\%\\
    2017 & 2018 & 26 & 100.0\% & 100.0\% & 100.0\%\\
    2018 & 2019 & 26 & 96.2\% & 100.0\% & 100.0\%\\
    2019 & 2020 & 28 & 96.4\% & 96.4\% & 96.4\%\\
    2020 & 2021 & 29 & 89.7\% & 89.7\% & 93.1\%\\
    \multicolumn{3}{|l}{\textit{Average}} & 94.0\% & 94.8\% & 95.5\%\\
    \multicolumn{6}{|l|}{\textit{5 year ahead}} \\
    2011 & 2016 & 16 & 75.0\% & 81.2\% & 87.5\%\\
    2016 & 2021 & 22 & 81.8\% & 86.4\% & 100.0\%\\
    \multicolumn{6}{|l|}{\textit{10 year ahead}} \\
    2011 & 2021 & 13 & 76.9\% & 84.6\% & 84.6\%\\
    \hline
    \end{tabular}
\end{table}
}

\setcitestyle{round}

\newcommand{\tlast}[0]{t^{\mathrm{last}}}

\newcommand{\tstart}[0]{t^{\mathrm{start}}}
\newcommand{\tend}[0]{t^{\mathrm{end}}}
\newcommand{\tpeak}[0]{t^{\mathrm{peak}}}

\title{Bayesian Projection of Extant Refugee and Asylum Seeker Populations}

\sectionfont{\fontsize{14}{15}\selectfont}
\subsectionfont{\fontsize{12}{15}\selectfont}

\begin{document}

\maketitle

\section*{Abstract}
Estimates of future migration patterns are of broad interest in demography. Forced migration, including refugee and asylum seekers, plays an important role in overall migration patterns, but is notoriously difficult to forecast. Focusing on refugees and asylum seekers, we propose a modeling pipeline based on Bayesian hierarchical time-series modeling for projecting refugee population official statistics by country of origin using data from the United Nations High Commissioner for Refugees (UNHCR). Our approach is based on a conceptual model of refugee and asylum seeker populations following growth and decline phases, separated by a peak. The growth and decline phases are modeled by logistic growth and decline through an \textit{interrupted logistic process model}. We evaluate our method through a set of validation exercises that show it has good performance for forecasts at 1, 5, and 10 year horizons, and we present projections for 35 countries of origin of large refugee and asylum seeker population.

\paragraph{Keywords} Bayesian hierarchical model $\cdot$ Refugees $\cdot$ Population projections

\section*{Introduction}

The study of refugee and asylum seeker populations is a large and evolving research area in demography \citep{arar2023refugee, graeme2018refugees}. Migration has traditionally been viewed as difficult to forecast due to inherent uncertainty in political and economic factors that are drivers of migration \citep{fuchs2021migrationforecasting, devalk2022migration}, and refugee and asylum seeker populations as a subset of international migration are particularly challenging to forecast. For example, the sudden outbreak of a war can lead very quickly to large numbers of refugees \citep{martin2018forecast}. Complicating matters further is that defining who qualifies as a refugee or asylum seeker is in itself a complex political and legal subject \citep{reed2016forcedmigration}. 

We propose an approach to projecting refugee and asylum seeker statistics by country of origin. Importantly, we focus on producing projections for countries with extant large reported populations of refugees and asylum seekers; we do not attempt to forecast which countries will begin to be new origins of refugees and asylum seekers. Our forecasting approach is based on a conceptual model of refugee and asylum seeker populations (stocks) in which these populations experience growth and decline phases, separated by a peak. The growth phase corresponds to refugees and asylum seekers leaving their country of origin due to a conflict or other reason. The peak is a period during which the refugees and asylum seeker stay in their country of destination,  or move to a alternate host country. During the decline stage, the refugees and asylum seekers return home or lose protected status for other reasons, such naturalization in their host country. 

Trends on either side of the peak are modeled as following logistic growth before the peak and decline after it. We refer to this process model as an \textit{interrupted logistic model}. We produce probabilistic projections of combined refugee and asylum seeker stocks using Bayesian statistical inference, drawing on the long line of previous research applying Bayesian methods to estimate and project demographic indicators \citep{bijak2011migration, bijak2016bayesiandemography,azose2015bayesianmigration,azose2016probabilistic}. 

The first step in modeling refugee and asylum seeker counts is to precisely define what is meant by these two terms. Other demographic indicators of interest, such as fertility or mortality rates, are defined by reference to immutable and universally recognized events. In this sense, the definition of what qualifies as a birth or a death is not controversial. The same definitional clarity is not manifest when it comes to tracking refugee and asylum seeker populations, as deciding who qualifies as a refugee or asylum seeker is in itself a complex political and legal subject that is continually evolving. In this work, we choose to define refugees and asylum seekers following definitions in the International Recommendations on Refugee Statistics \citepalias{expert2018international}. This choice is in part practical, in that the UN High Commissioner for Refugees (UNHCR) releases official statistics of refugee and asylum seeker populations based on these definitions \citep{unhcrdatabase}.  We emphasize that our projections of refugee and asylum seeker populations are therefore to be solely understood as projections of refugee and asylum seeker populations as those populations are defined by the EGRIS recommendations and reflected in UNHCR statistics.

Long-term forecasts of refugee populations have been produced as inputs to world population projections; in particular, the United Nations (UN) Population Division is a major producer of population projections through regular releases of the World Population Prospects. The UN uses probabilistic Bayesian approaches for projecting total fertility and life expectancy \citep{alkema2011tfr, raftery2013life,raftery2014}. For refugees, UN projections are currently based on an assumption that two thirds of refugees will return to their country of origin within 5 years \citepalias{un2022wpp}. We adopt this simple method as a benchmark against which to compare our proposed approach. 

Other attempts have been made to apply techniques from statistical demography to project forcibly displaced populations. Within the UN High Commission for Refugees (UNHCR) there have been multiple such efforts. The UNHCR Demographic Projection Tool applies demographic methods to project age-specific refugee populations based on fertility and life expectency data. However, the tool does not attempt to project arrivals or departures of refugees, rather relying on the analyst to provide possible future scenarios. From another angle, the UNHCR Jetson project seeks to apply predictive modeling to forecast refugee movements \citepalias{jetson2024}. A predictive model of monthly arrivals of internally displaced persons by region in Somalia was developed using climate, weather, and market variables as predictors \citepalias{jetson2024somalia}. 

Statistical and machine learning approaches have been used to predict refugee and asylum seeker movements using indicators such as violent conflicts, market prices, and weather and climate variables  \citep{carammia2022asylum, huynh2020idps, singh2019bayesian}. Typically, such methods have shorter forecast horizons (on the order of weeks or months) and are targeted towards aiding in humanitarian response efforts. Similarly, a number of countries have implemented prediction systems in order to anticipate arrivals of asylum seekers \citep{angenendt2023}. Methods based on detecting signals in other variables are difficult to apply for medium- to long-term population projections, as doing so typically requires projecting the indicators themselves, which are often noisier than the demographic outcome being forecast. As such, our approach is based only on historical refugee data. A possible downside of this approach is that our methods may be less well suited for short-term projections of the type useful for humanitarian planning in emergencies.

Gravity-type models have also been applied to predict forced migration flows \citep{qi2023forcedmigration, saldarriaga2019gravity}. However, the fundamental assumptions of gravity models have been criticized as insufficient in explaining human migration patterns \citep{beyer2022gravity,welch2022bilateral}. In addition, similarly to the approaches described in the previous paragraph, gravity models typically depend on complex covariates including economic, climate, and conflict variables. Projections based on gravity models therefore require projections of all covariates, which is difficult especially for longer-term projections.

A related methodological area of research has been the development of agent-based models (ABMs) to simulate forced migration on a variety of geographic and temporal time scales \citep{gray2017choosing,sokolowski2014abm,groen2016abm,frydenlund2018abm,kniveton2011abm}. Typically ABMs simulate individual refugees according to a set of complex behavioral rules. ABMs may be developed for reasons other than forecasting, such as simulating the effect of interventions on outcomes of interest. The intensive computational requirements and unclear probabilistic properties of ABMs complicates their application to producing longer-term forecasts, especially when statistically well-founded estimates of uncertainty are of interest. Recent work on ABM simulations of migration incorporating uncertainty via Bayesian principles provides a potential path forward \citep{bijak2021}.

The rest of the paper unfolds as follows. In the next section we describe the analysis dataset and propose the statistical methodology for our approach. Out-of-sample validations and substantive results are presented in the following section. We conclude with a discussion of the findings in the final section.

\section*{Methods}
\label{section:methods}

\subsection*{Data}
We focus on individuals classified as refugees and asylum seekers by UNHCR based on definitions in the International Recommendations on Refugee Statistics \citepalias{expert2018international}. 
The number of refugees and asylum seekers by country of origin were sourced from the UNHCR Refugee Population Statistics Database \citep{unhcrdatabase}. Refugees are those ``persons who have current refugee status, granted either before arrival or upon arrival
in the receiving country" (\citetalias{expert2018international}; see within item 92.3 for extended definition, and Chapter 2 for legal discussion of refugee status). Asylum seekers are defined as ``Persons who have filed an application for asylum in a country other than their own and
whose claims have not yet been determined. These include those filing primary applications or subsequent
applications following an appeal. The date on which the application for asylum is filed marks their entry into
the status of asylum seeker. They remain in the status of asylum seeker until their application is considered
and adjudicated." \citepalias[92.2]{expert2018international}

The UNHCR releases population statistics twice per year, with end of year statistics being released the following June. Based on definitions provided in the International Recommendations on Refugee Statistics, forcibly displaced and stateless populations are separated into five categories: refugees, asylum seekers, internally displaced persons of concern to UNHCR, other people in need of international protection, and stateless people \citepalias{expert2018international}. Statistics on each population type are available starting in different years depending on the host UN division.

Formally, let $t = 1, \dots, T$ index years and $c = 1, \dots, C$ index UN divisions (henceforth referred to generically as ``countries"). The UN divisions ``Unknown", ``Stateless", ``Western Sahara", and ``Palestinian" were excluded from the analysis -- Western Sahara for data quality concerns \citepalias{ussd2007}, and Palestinian because those data are sourced differently than other divisions \citep{unhcrdatabase}. Let $R_{c,t}$ be the combined number of refugees and asylum seekers with origin country $c$ at time $t$. Let $P_{c,t}$ be the population of country $c$ at time $t$, using as source the UN World Population Prospects 2022 \citepalias{wpp2022}. We model refugee and asylum seeker populations as proportions, defining $r_{c,t} = R_{c,t} \slash \left(R_{c,t} + P_{c,t}\right)$. The analysis dataset was formed by selecting countries with at least 10 years of observed data, at least one observed proportion $r_{c,t}$ greater than 1\%, and at least one observed population size $R_{c,t}$ greater than 1000.

\subsection*{Conceptual Framework}
\label{section:framework}
We conceptualize the number of refugees and asylum seekers from a particular country over time as arising from a succession of one or more periods characterized by growth and subsequent decline which we refer to generically as a \textit{growth/decline period} or, succinctly, as an \textit{episode}. 
We think of an episode as being caused by ongoing events or conditions in the country, such as war, famine, or political/economic upheaval. The refugee and asylum seeker population is expected to increase while the condition is present, before declining after the condition ends. An episode thus comprises two stages: an initial growth stage and a decline phase, separated by a peak. We note that a notable limitation of our conceptual framework is that it may not well capture long-term refugee populations in which population dynamics are dominated by traditional demographic forces.

Figure~\ref{fig:pipeline} shows illustrative data from three countries. In Mozambique, a driver of the refugee and asylum seeker population was the Mozambican Civil War from 1977-1992. Thus, we see a sharp increase in the refugee and asylum seeker population during the war, a peak in 1992 when the war ended, followed by a decline. Somalia appears to have experienced two episodes, with the most recent episode currently in the decline phase. Honduras appears to currently be in the growth phase of a episode, and the timing and magnitude of the peak of the episode is not yet known.

\pipelinefigure

\subsection*{Modeling Approach}
Our modeling approach is informed by the conceptual framework for episodes. First, refugee and asylum seeker populations from each origin country are divided into growth/decline periods. Next, a Bayesian hierarchical model is fit based on the segmented dataset. The overall pipeline is illustrated in Figure~\ref{fig:pipeline} using Mozambique, Somalia, and Honduras as examples.

\label{section:approach}
\paragraph{Segmentation} The time series for each country is first segmented into one or more episodes of growth and decline. We do this by identifying local peaks in the time series, which we take to be the peak of a corresponding episode. To identify local peaks we use a peak detection algorithm described by \cite{palshikar2009peakdetection} and implemented in the \texttt{scorepeak} package \citep{ochi2019scorepeak}. Once the peaks are identified, we identify the extent of the episode by looking for troughs before and after the peak. A \textit{trough} is defined as the minimum proportion observed between two peaks (or between a peak and the beginning or end of the time series). The start of a episode is defined as the first year after a trough (or the beginning of the time-series) in which the refugee and asylum seeker proportion $r_{c,t}$ is greater than 0.025\%. If such a year does not exist, the start is considered to be the year before the peak.

Each episode is next classified as ended or ongoing. An episode is classified as ended if it is either followed by another episode, or if the most recent observed proportion $r_{c,t}$ is less than 0.025\%. For an ended episode, the end time point is set to the year in which the last observation was greater than 0.025\%. All other episodes are classified as ongoing. Ongoing episodes are further divided into being in the growth or decline phase.  Countries that have a detected local peak on or after 2015, or that have a proportion in 2021 greater than 80\% of the most recent peak, are classified as being in the growth phase. All other ongoing episodes are classified as being in the decline phase.

Figure~\ref{fig:pipeline} illustrates the detected episodes for Mozambique, Somalia, and Honduras. Two episodes are detected in Mozambique, corresponding roughly to the Mozambican War of Independence and the Mozambican Civil War. The second episode in Mozambique is classified as having ended, as all observations after 1995 are under 0.025\%. Somalia also has two episodes detected, corresponding to different phases of the Somali Civil War. The second episode in Somalia is classified as ongoing and in the decline phase, as the most recent observations are all over 0.025\%. Finally, an ongoing episode is also detected in Honduras, but is classified as being in the growth phase. Table~\ref{tab:country_crisis_classifications} shows the classifications of the final episode in each country in the analysis dataset.

\classificationtable

Formally, let $m = 1, \dots, M$ index the detected episodes from the segmentation step, and denote by $c[m]$ the country corresponding to the $m$th episode. Let $\tstart_m$, $\tend_m$ be the start and end time points of the $m$th episode, respectively. Let $\tpeak_m$ be the time point of the peak of episode $m$. 

\paragraph{Process Model}
Let $\mu_{c,t}$ denote the modeled refugee proportion in country $c$ at time $t$. The process model describes how $\mu_{c,t}$ is expected to evolve over time \citep{susmann2022tmmps}. The core of our modeling approach is to assume that refugee and asylum seeker populations follow logistic growth and decline trends during the growth and decline phases of a episode, respectively. The rate of change of $\mu_{c[m], t}$ during the growth and decline phases is modeled by the following logistic rate function:
\begin{align}
    f(\mu_{c[m], t}, \omega, \lambda) = \begin{cases}
        \lambda - \mu_{c[m],t}, & \mu_{c[m],t} > \lambda, \\
        \omega \cdot \mu_{c[m],t} \left(1 - \frac{\mu_{c[m],t}}{\lambda} \right), & \text{otherwise.}
    \end{cases}
\end{align}
where $\omega \in \mathbb{R}$ is a rate parameter and $\lambda \in (0, 1)$ is an asymptote parameter. During the growth phase, an episode-specific rate parameter $\omega_m^{(1)} > 0$ is applied. During the decline phase a rate parameter $\omega_m^{(2)} < 0$ is used. Both growth and decline phases share an asymptote parameter $\lambda_m$. Formally, the expected rate of change for episode $m$ at time $t$ is given by
\begin{align}
    \xi_{m,t} = \begin{cases}
        f(r_{c[m],t-1}, \omega_m^{(1)}, \lambda_m), &  \tstart_m \leq t \leq \tpeak_m, \\
        f(r_{c[m],t-1}, \omega_m^{(2)}, \lambda_m), & \tpeak_m < t \leq \tend_m.
    \end{cases}
\end{align}
We refer to the process model as an \textit{interrupted logistic process model} because the logistic growth during the growth phase is interrupted at time $\tpeak_m$. Figure~\ref{fig:logistic_model} illustrates the parameters of the interrupted logistic model for a single episode.

\logisticmodelfigure

\paragraph{Data Model}
We assume that the observed rate of change is normally distributed around the expected rate of change, with country-specific variance terms that differ for the growth and decline phases:
\begin{align}
    r_{c[m],t} - r_{c[m],t-1} \sim N(\xi_{m,t}, \sigma^2_{c,t}),
\end{align}
where
\begin{align}
    \sigma_{c,t} = \begin{cases}
        \sigma_{c}^{(1)}, & t \leq \tpeak_m, \\
        \sigma_{c}^{(2)}, & t > \tpeak_m.
    \end{cases}
\end{align}

\paragraph{Episode lengths and maxima}
We model the length of the growth phase and the refugee and asylum seeker proportion observed at the peak for use in projecting episodes that are still in the growth phase. The length of the growth phase is assumed to be exponentially distributed with rate parameter $\psi > 0$:
\begin{align}
    \tpeak_m - \tstart_m \sim \mathrm{Exp}(\psi),
\end{align}
with prior $\psi\sim \mathrm{InvGamma}(0.1, 0.1)$, an inverse Gamma distribution.
The exponential model was chosen based on empirical inspection of the distribution of growth phase lengths (Appendix Figure~\ref{fig:growth-phase-lengths}); we note also the natural connection between the exponential distribution and the distribution of waiting times.
The proportion at the peak is assumed to follow a logit-normal distribution around a mean parameter $\rho$:
\begin{align}
    \mathrm{logit}(r_{c[m], \tpeak_m}) \sim N(\rho, \sigma_\rho^2),
\end{align}
where $\mathrm{logit}(x) = \log(x/ (1 - x))$. 
The following priors are assigned to $\rho$ and $\sigma_{\rho}$:
\begin{align}
    \rho        &\sim N(0, 3), \\
    \sigma_\rho &\sim \mathrm{InvGamma}(0.1, 0.1). \\
\end{align}

\paragraph{Hierarchical Priors}
Hierarchical priors are used to share information between episodes. The upper asymptote of each episode is assumed to follow a logit-normal distribution around an overall mean:
\begin{align}
    \mathrm{logit}(\lambda_m) \sim N(\lambda_g, \sigma_\lambda^2).
\end{align}
The growth and decline rate are assumed to follow a Student's $t$-distribution around an overall mean:
\begin{align}
    \omega_m^{(1)} &\sim t_3(\mu_{\omega^{(1)}}, \sigma_{\omega^{(1)}}^2), \\
    \omega^{(2)}_m &\sim t_3(\mu_{\omega^{(2)}}, \sigma_{\omega^{(2)}}^2).
\end{align}
where $t_\nu(\mu, \sigma^2)$ denotes a non-centered Student's $t$-distribution with $\nu$ degrees of freedom, location $\mu$ and scale $\sigma^2$. The Student's $t$-distribution was chosen to allow for outlier growth and decline rates. 
Similar hierarchical priors are used for the growth and decline noise parameters:
\begin{align}
    \log \sigma_c^{(1)} &\sim N(\mu_{\sigma^{(1)}}, \sigma_{\sigma^{(1)}}^2), \\
    \log \sigma_c^{(2)} &\sim N(\mu_{\sigma^{(2)}}, \sigma_{\sigma^{(2)}}^2). \\
\end{align}
The priors used for the hierarchical models are as follows, for the growth and decline rates:
\begin{align}
    \mu_{\omega^{(1)}} &\sim N(0, 3), \\
    \mu_{\omega^{(2)}} &\sim N(0, 3), \\
    \sigma_{\omega^{(1)}} &\sim \mathrm{InvGamma}(0.1, 0.1), \\
    \sigma_{\omega^{(2)}} &\sim \mathrm{InvGamma}(0.1, 0.1),
\end{align}
and for the growth and decline noise parameters:
\begin{align}
    \mu_{\sigma^{(1)}} &\sim N(-7, 4), \\
    \mu_{\sigma^{(2)}} &\sim N(-7, 4), \\
    \sigma_{\sigma^{(1)}} &\sim \mathrm{InvGamma}(0.1, 0.1) \\
    \sigma_{\sigma^{(2)}} &\sim \mathrm{InvGamma}(0.1, 0.1).
\end{align}

\subsection*{Projections}
Projections are produced for all ongoing episodes. Suppose we have at our disposal $k = 1, \dots, K$ posterior draws from the joint posterior distribution of the model parameters conditional on all the observed data. We will append an additional suffix to denote the $k$th posterior draw, as in $\alpha_{k}$ to denote the $k$th posterior draw of the parameter $\alpha$. 

The projection method is different depending on whether the ongoing episode was classified as being in the decline or growth phase. For a episode $m$ that is ongoing and in the decline phase, the timing and level of the peak is taken as fixed. Thus, for all $t > \tlast_m$, the refugee and asylum seeker proportion is projected by recursive application of the following equation for each posterior draw $k = 1, \dots, K$:
\begin{align}
    \mu_{c[m], t, k} \sim N\left(\mu_{c[m], t-1, k} + f\left(\mu_{c[m], t-1, k}, \omega_{m,k}^{(2)}, \lambda_{m,k}\right), \sigma^{2}_{c,t,k}\right),
\end{align}
and where we set the initial value $\mu_{c[m], \tlast_m, k} = r_{c[m], \tlast_m}$. If, for any time $t$, $\mu_{c[m], t,k} < 0$, then we set $\mu_{c[m], t,k} = 0.001$ before continuing recursive application of the formula. 

For an ongoing episode $m$ in the growth phase, the timing and level of the peak has not been observed, and we integrate over this uncertainty. Suppose the peak falls in year $\tlast_m + \delta$, where $\delta \in \mathbb{N}^+$ (in practice we take $\delta \in \{ 0, 1, \dots, 15 \})$. Conditional on $\delta$, projections are generated by recursive application of the equation
\begin{align}
    \mu_{c[m], t+1,\delta,k} \sim \begin{cases}
        N\left(\mu_{c[m], t,\delta,k} + f\left(\mu_{c[m], t,\delta,k}, \omega_{m,k}^{(1)} \right), \lambda_{m,k}, \sigma^{2}_{c,t,k} \right), & t \leq \tlast_m + \delta, \\
        N\left(\mu_{c[m], t,\delta, k} + f\left(\mu_{c[m], t,\delta,k}, \omega_{m,k}^{(2)} \right), \lambda_{m,k}, \sigma^2_{c,t,k} \right), & t > \tlast_m + \delta.
    \end{cases}
\end{align}
As before, if for any $t$ $\mu_{c[m], t, \delta,k} < 0$, then we set $\mu_{c[m], t, \delta,k} = 0.001$.
Next, the probability of $\delta$ conditional on the implied episode length and the refugee and asylum seeker proportion at the peak is calculated conditional on the episode length parameters $\psi$ and peak proportion parameters $\rho$ and $\sigma_\rho^2$ previously estimated.
\begin{align}
    p_\delta \propto f_{\mathrm{Exp}}(\delta \mid \psi_k) f_{\mathrm{N}}(\mathrm{logit}(\mu_{c[m], \tlast_m + \delta, \delta, k}) \mid \rho, \sigma^2_\rho),
\end{align}
where $f_\mathrm{Exp}$ is the density of the Exponential distribution and $f_N$ the density of the normal distribution. 
A final peak position $\delta^*$ is drawn with probabilities $p_{\delta}$, and the final projections are set to $\mu_{c[m], t, k} = \mu_{c[m], t, \delta^*, k}$. Figure~\ref{fig:interrupted-logistic-ongoing} illustrates how projections for ongoing growth episodes are generated conditional on $\delta$. 

\interruptedlogisticongoingfigure

\subsection*{Benchmark}
As a benchmark we implemented a deterministic projection method based on the United Nations current rule of thumb that assumes two thirds of refugees will return to their country of origin within 5 years \citepalias{un2022wpp}. We assume that the last observed value is at the midpoint of a logistic decline curve. Taken together, these assumptions imply a logistic rate of change of
\begin{align}
    \omega^* = -\frac{1}{5} \log 2.
\end{align}
Projections for any year $t$ in a country $c$ with final observed value at $\tlast_c$ are then given by
\begin{align}
    \mu_{c,t}^* = \frac{2 r_{c,\tlast_c}}{1 + \exp(\omega^*(t - \tlast_c)}. 
\end{align}

\section*{Results}
\label{section:results}

\subsection*{Validation}
\label{section:validation}
As a validation exercise, we held out all observations after a cutoff time point $t^*$ and applied the full projection pipeline. The validation set included all countries that did not have a new episode start after the cutoff time point. 

The mean error and mean absolute error for the proposed Bayesian method and the benchmark deterministic logistic method were computed for $1$-year, $5$-year, and $10$-year projection horizons. The results are presented in Table~\ref{tab:validation_error}. For the $1$-year ahead projections, the Bayesian method had lower mean error and mean absolute error than the benchmark. The benchmark performed slightly better in terms of mean error for the $5$- and $10$-year projections than the Bayesian method.

For the Bayesian method  we computed the 80\%, 90\%, and 95\% empirical coverage rates of the corresponding credible intervals. The results are shown in Table~\ref{tab:validation_coverage}. The small sizes of the validation sets make it difficult to draw strong conclusions as to the quality of the model calibration. For the $1$-year ahead projections, the credible intervals are generally conservative, with the 80\% intervals tending to have higher than nominal empirical coverage. The longer-term projections exhibit reasonable performance, with for example the $5$-year ahead projections from 2016 having near-nominal $77.3\%$, $86.4\%$, and $95.5\%$ empirical coverage for the 80\%, 90\%, and 95\% credible intervals, respectively.

\errortable

\coveragetable

\subsection*{Projections}
\label{section:projections}
Projections for all countries with an ongoing episode detected as of 2021 are included in the appendix. Illustrative projections for six of these countries are shown in Figure~\ref{fig:example-projections}. Of these countries, Colombia, Gambia, and Somalia have episode in the decline phase, and El Salvador, Haiti, and Honduras have episodes in the growth phase. The strength of the probablistic projections from the  Bayesian model as compared to the deterministic benchmark can be seen in Honduras. Despite the posterior median projection being similar to the deterministic projection, the Bayesian credible intervals indicate the possibility of continued growth in the refugee and asylum seeker population. In Gambia, the posterior median projects faster decline than the deterministic benchmark because the Bayesian model incorporates the previous rates of decline observed in the episode.

\exampleprojectionsfigure

Figure~\ref{fig:projection-el-salvador} illustrates how projections for El Salvador would have changed year over year as new data became available. In 2016, the 95\% credible intervals include the possibility of an increase in the refugee and asylum seeker population to around the same level as the peak of the previous episode. For the 2017 estimates, the model adjusts upwards its projections based on the increased rate of change implied by the new data point. By 2021, the posterior median suggests the episode has reached its peak, although there is still significant posterior mass on the possibility of continued increases up a level last seen in the first episode in the country in the 1980s.

\exampleprojectionfigure

\section*{Discussion}
\label{section:discussion}

We have proposed a Bayesian hierarchical time-series model of refugee and asylum seeker population (stocks) by country of origin based on UNHCR data. We then used the model to project future refugee and asylum seeker population statistics for countries of origin of large refugee and asylum seeker populations. Validation exercises suggest the model is well-calibrated at multiple time horizons. 

UN population projections assume that two thirds of refugees will return to their country of origin within 5 years \citepalias{un2022wpp}. Our validation results show that this rule of thumb performs well for 5-year ahead projections. The Bayesian modeling approach achieves comparable or better performance in all the validation exercises, and has the additional benefit of providing well-calibrated probabilistic projections. 

A fundamental limitation of our projections is that they are generated based on official statistics gathered and disseminated by the UNHCR, using their definition of refugees and asylum seekers and based on their data sources and data collection methods. However, constructivist legal definitions of refugee and asylum seeker categories exclude populations who may fall into those same categories in a broader sociological sense. Furthermore, adjudication of refugee and asylum seeker status are inherently political \citep{fitzgerald2018refugee}. The UNHCR may, for example, end recognition of refugee status based on a determination of improved circumstances in a country or origin \citep{fitzpatrick2003cessation}. Affected individuals would no longer be counted in official statistics, even if they would still be considered a refugee under alternative definitions. The inherent limitations of the data imply that our projections can only be interpreted as projections of refugee and asylum seeker \textit{official statistics}, rather than projections of refugee and asylum seeker populations in a broader sense. While our modeling approach is likely applicable to the latter task, generating projections would require alternative data sources. 

A basic assumption of any statistical forecasting method is that the future will be in some way similar to the past. In this work, this assumption manifests itself in the model specification that assumes current growth and decline periods will follow a similar shape to previous episodes. However, the underlying dynamics driving refugee and asylums seeker dynamics may change in the future in ways that fundamentally change refugee population trends. Global climate change is a prominent example of a factor that may drive future displacement in ways that differ from past causes of displacement \citep{piguet2011}. Uncertainty estimates from our modeling approach, however, are conditional on the model specification and do not incorporate uncertainty in how displacement dynamics may change in the future. As such, while credible intervals from our approach appear to be relatively well-calibrated in the projection horizons considered in the validation exercises, there is no guarantee that the model specification will lead to good performance in the long term.

Our work leaves open multiple directions for future research. We focused on projecting combined refugee and asylum seeker populations based on definitions adopted by the UNHCR and reflected in official statistics; it may be of interest to examine alternative subsets of forced migrants and using other data sources. Our modeling method is adaptable to consider other populations that follow dynamics captured by our conceptual model.  An important limitation of our work is that it focuses only on forecasting how current refugee and asylum seeker counts will evolve, and, hopefully, end. We did not attempt to forecast the start of large new refugee or asylum seeker populations, rather focusing on projecting how existing large populations will change over time. This focus encodes an optimistic assumption that, in the long term, all refugee and asylum seeker populations will go to zero by those affected returning home or losing status in other ways, such as naturalisation. However, research has been carried out on identifying where and when future episodes will occur \citep{oecd2019earlywarning}. Producing accurate ``early warnings" is notoriously difficult due to the complexity of causes of refugee migration \citep{schmeidl1996earlywarning}, although progress has been made \citep{napierala2022earlywarning}. An area for future research is to expand our approach to incorporate uncertainty about whether new crises will arise in the future. In addition, incorporating estimates of refugee and asylum seeker statistics in country-level population projections requires forecasting in which destination countries refugee and asylum seekers will seek protection. To address this in the future, it could be useful to draw on developments in projecting bilateral migration flows between countries \citep{welch2022bilateral}. 

\ifx\arxiv\undefined 
\else
\paragraph{Acknowledgements:} This research was supported by NIH grant R01 HD070936. The authors thank Patrick Gerland for helpful conversations. Both authors thanks the Laboratoire MAP5 at Universit\'{e} Paris-Cit\'{e} for warm hospitality.
\fi

\bibliography{bibliography}

\pagebreak

\section*{Appendix}

\setcounter{figure}{0}
\renewcommand{\thefigure}{A\arabic{figure}}

\subsection*{Observed episode lengths}
\begin{figure}[h]
    \centering
    \includegraphics[width=0.9\textwidth]{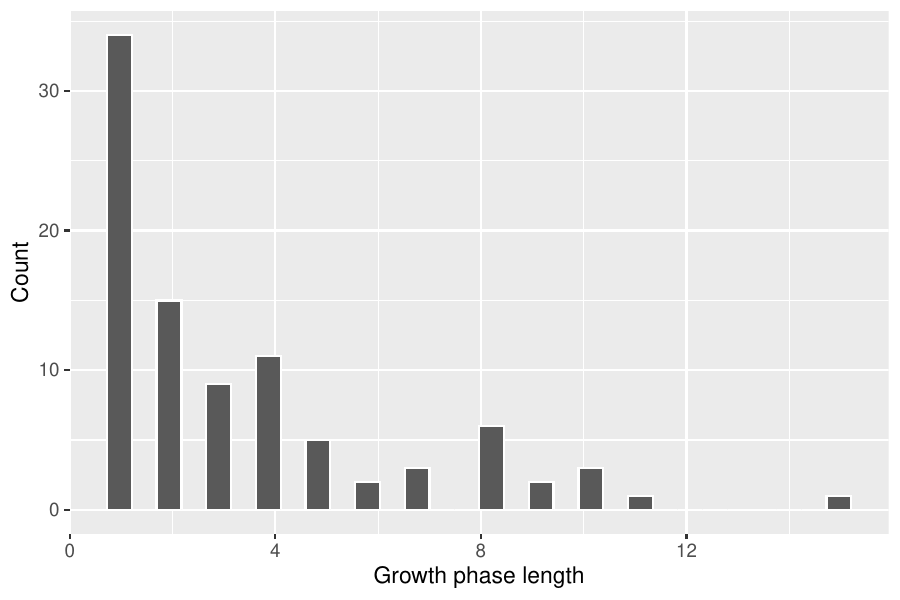}
    \caption{Empirical distribution of the length of the growth phase for all episodes classified as ended.}
    \label{fig:growth-phase-lengths}
\end{figure}

\subsection*{Sensitivity analysis}

We conducted a sensitivity analysis on the threshold for defining an episode. In the main text we consider a threshold of 0.025\%. As a sensitivity analysis, we conducted the complete analysis using thresholds of 0.0125\% and 0.05\%. 

At a threshold of 0.0125\%, 9 additional episodes were detected compared to using the 0.025\% threshold. At a threshold of 0.05\%, 7 fewer episodes were detected compared to using the 0.025\% threshold. Table~\ref{tab:validation_error_sensitivity} shows mean error and mean absolute error for the Bayesian and benchmark logistic method for both thresholds, and Table~\ref{tab:validation_coverage_sensitivity} shows empirical coverage results. The Bayesian method typically had lower MAE and ME for both alternative thresholds compared to the benchmark method, particularly in the 1-year ahead validations. 

\begin{table}[H]
    \centering
    \caption{\label{tab:validation_error_sensitivity} Mean Absolute Error (MAE) and Mean Error (ME) of posterior median projections in the validation exercises.}
    \begin{tabular}{|llrrrrr|}
    \hline
    & & & \multicolumn{2}{c}{MAE} & \multicolumn{2}{c|}{ME} \\
    Cutoff & Target & $n$ & Bayes & Logistic & Bayes & Logistic \\
    \hline
    \multicolumn{7}{|l|}{\textbf{Threshold 0.00125\%}} \\
    \multicolumn{7}{|l|}{\textit{1 year ahead}} \\
    2016 & 2017 & 33 & \textbf{0.55} & 0.64 & \textbf{0.47} & 0.59\\
    2017 & 2018 & 35 & \textbf{0.15} & 0.21 & \textbf{0.03} & 0.18\\
    2018 & 2019 & 35 & \textbf{0.15} & 0.19 & \textbf{0.03} & 0.18\\
    2019 & 2020 & 36 & \textbf{0.14} & 0.24 & \textbf{0.04} & 0.20\\
    2020 & 2021 & 37 & \textbf{0.23} & 0.35 & \textbf{0.15} & 0.30\\
    \multicolumn{7}{|l|}{\textit{5 year ahead}} \\
    2011 & 2016 & 19 & \textbf{1.25} & 1.35 & 0.76 & \textbf{0.68}\\
    2016 & 2021 & 30 & 1.40 & \textbf{1.34} & \textbf{1.25} & 1.31\\
    \multicolumn{7}{|l|}{\textit{10 year ahead}} \\
    2011 & 2021 & 16 & \textbf{1.87} & 1.89 & 1.56 & \textbf{1.48} \\
    \multicolumn{7}{|l|}{\textbf{Threshold 0.5\%}} \\
    \multicolumn{7}{|l|}{\textit{1 year ahead}} \\
    2016 & 2017 & 23 & \textbf{0.77} & 0.88 & \textbf{0.65} & 0.80\\
    2017 & 2018 & 23 & \textbf{0.23} & 0.31 & \textbf{0.06} & 0.27\\
    2018 & 2019 & 24 & \textbf{0.19} & 0.27 & \textbf{0.03} & 0.25\\
    2019 & 2020 & 25 & \textbf{0.17} & 0.30 & \textbf{0.02} & 0.25\\
    2020 & 2021 & 27 & \textbf{0.30} & 0.47 & \textbf{0.20} & 0.40\\
    \multicolumn{7}{|l|}{\textit{5 year ahead}} \\
    2011 & 2016 & 12 & \textbf{1.93} & 2.09 & 1.16 & \textbf{1.06}\\
    2016 & 2021 & 20 & 2.01 & \textbf{1.91} & \textbf{1.81} & 1.85\\
    \multicolumn{7}{|l|}{\textit{10 year ahead}} \\
    2011 & 2021 & 10 & \textbf{2.91} & 2.96 & 2.43 & \textbf{2.31}\\
    \hline
    \end{tabular}
\end{table}

\begin{table}[H]
    \centering
    \caption{Empirical coverage of 80\%, 90\%, and 95\% credible intervals for projections in the validation exercises for additional values of the episode threshold.}
    \label{tab:validation_coverage_sensitivity}
    \begin{tabular}{|llrrrr|}
    \hline
    & & & \multicolumn{3}{c|}{Coverage} \\
    Cutoff & Target & $n$ & 80\% & 90\% & 95\% \\
    \hline
    \multicolumn{6}{|l|}{\textbf{Threshold 0.025\%}} \\
    \multicolumn{6}{|l|}{\textit{1 year ahead}} \\
    2016 & 2017 & 33 & 87.9\% & 90.9\% & 90.9\%\\
    2017 & 2018 & 35 & 100.0\% & 100.0\% & 100.0\%\\
    2018 & 2019 & 35 & 94.3\% & 97.1\% & 100.0\%\\
    2019 & 2020 & 36 & 94.4\% & 94.4\% & 97.2\%\\
    2020 & 2021 & 37 & 86.5\% & 94.6\% & 97.3\%\\
    \multicolumn{3}{|l}{\textit{Average}} & 92.6\% & 95.5\% & 97.2\%\\
    \multicolumn{6}{|l|}{\textit{5 year ahead}} \\
    2011 & 2016 & 19 & 73.7\% & 78.9\% & 89.5\%\\
    2016 & 2021 & 30 & 83.3\% & 90.0\% & 96.7\%\\
    \multicolumn{6}{|l|}{\textit{10 year ahead}} \\
    2011 & 2021 & 16 & 62.5\% & 68.8\% & 75.0\%\\
    \multicolumn{6}{|l|}{\textbf{Threshold 0.05\%}} \\
    \multicolumn{6}{|l|}{\textit{1 year ahead}} \\
    2016 & 2017 & 23 & 87.0\% & 87.0\% & 87.0\%\\
    2017 & 2018 & 23 & 100.0\% & 100.0\% & 100.0\%\\
    2018 & 2019 & 24 & 95.8\% & 95.8\% & 100.0\%\\
    2019 & 2020 & 25 & 96.0\% & 96.0\% & 100.0\%\\
    2020 & 2021 & 27 & 85.2\% & 85.2\% & 88.9\%\\
    \multicolumn{3}{|l}{\textit{Average}} & 92.6\% & 92.6\% & 95.1\%\\
    \multicolumn{6}{|l|}{\textit{5 year ahead}} \\
    2011 & 2016 & 12 & 66.7\% & 75.0\% & 83.3\%\\
    2016 & 2021 & 20 & 85.0\% & 85.0\% & 100.0\%\\
    \multicolumn{6}{|l|}{\textit{10 year ahead}} \\
    2011 & 2021 & 10 & 70.0\% & 80.0\% & 80.0\%\\
    \hline
    \end{tabular}
\end{table}

\subsection*{Ongoing episode predictions}
\begin{figure}[H]
    \centering
    \caption{Posterior predictive distributions for the proportions $\mu_{m,t}$ for all episodes and time points. To simplify presentation, multiple episodes from the same country are shown in the same plot.}
    \label{fig:posterior-predictive-all-plots}
\end{figure}
\includepdf[pages={1-}, width=0.5\textwidth, nup=2x4]{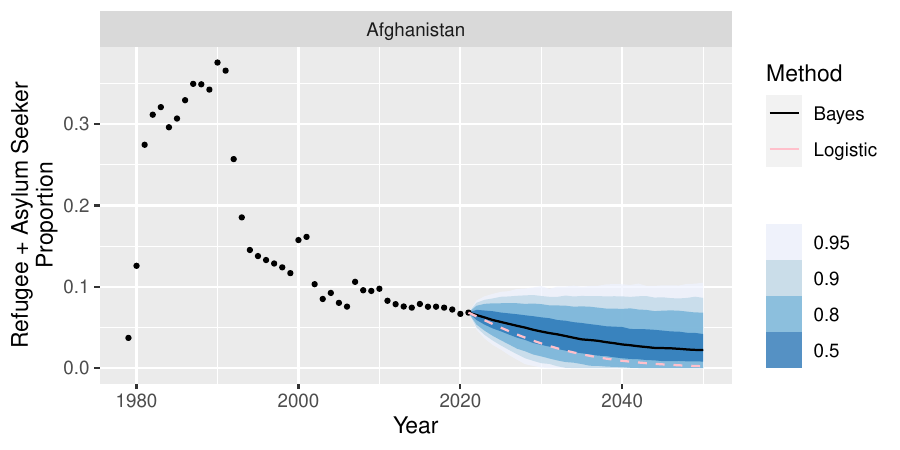}

\subsection*{All detected episodes}
\begin{figure}[H]
    \centering
    \caption{Detected episodes in each country according to the deterministic rules described in the main text (using the the threshold of 0.025\% as in the primary analysis). Each episode is further annotated with the detected growth and (when applicable) decline phases and peaks.}
    \label{fig:all-detected-episodes}
\end{figure}
\includepdf[pages={1-}, width=0.5\textwidth, nup=2x4]{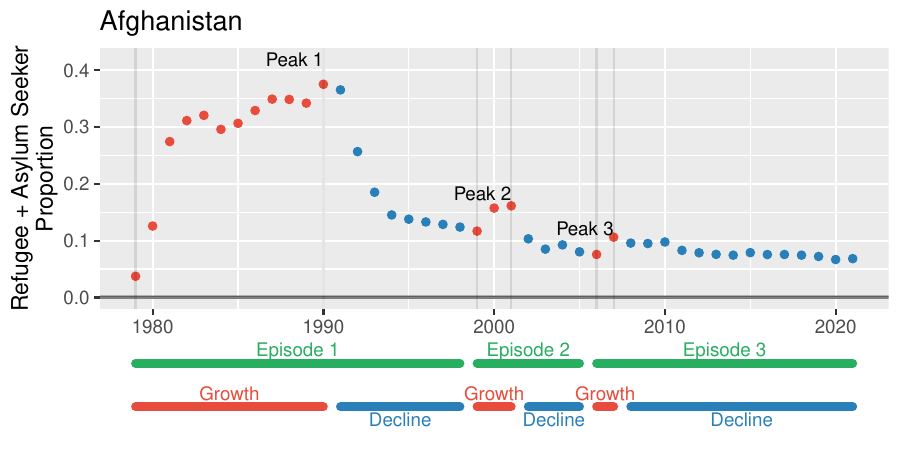}

\end{document}